\documentclass[%
 reprint,
 superscriptaddress,
 amsmath,amssymb,
 aps,
 prd,
 floatfix, 
]{revtex4-1}

\usepackage{textcomp}
\usepackage{gensymb}
\usepackage{graphics}
\usepackage{graphicx}
\usepackage{dcolumn}
\usepackage{multirow}
\usepackage{bm}
\usepackage{enumerate}
\usepackage[usenames,dvipsnames,svgnames,table]{xcolor}
\usepackage{hyperref}
\hypersetup{
	colorlinks=true,
	linktocpage=true,
	citecolor=YellowOrange,
	linkcolor=violet,
	urlcolor=MidnightBlue!50!Aquamarine,
}

\newcommand{\hc}{\text{h.c.}}
\newcommand{\brabar}[1]{\overset{\scalebox{.3}{(}\raisebox{-1.7pt}[0pt][0pt]{$-$}\scalebox{.3}{)}}{#1}}

\newcommand{\changes}[1]{\textcolor{black}{#1}}

\begin{document}

\title{Search for heavy neutrinos with the T2K near detector ND280}

\newcommand{\INSTHD}{\affiliation{University Autonoma Madrid, Department of Theoretical Physics, 28049 Madrid, Spain}}
\newcommand{\INSTEE}{\affiliation{University of Bern, Albert Einstein Center for Fundamental Physics, Laboratory for High Energy Physics (LHEP), Bern, Switzerland}}
\newcommand{\INSTFE}{\affiliation{Boston University, Department of Physics, Boston, Massachusetts, U.S.A.}}
\newcommand{\INSTD}{\affiliation{University of British Columbia, Department of Physics and Astronomy, Vancouver, British Columbia, Canada}}
\newcommand{\INSTGA}{\affiliation{University of California, Irvine, Department of Physics and Astronomy, Irvine, California, U.S.A.}}
\newcommand{\INSTI}{\affiliation{IRFU, CEA Saclay, Gif-sur-Yvette, France}}
\newcommand{\INSTGB}{\affiliation{University of Colorado at Boulder, Department of Physics, Boulder, Colorado, U.S.A.}}
\newcommand{\INSTFG}{\affiliation{Colorado State University, Department of Physics, Fort Collins, Colorado, U.S.A.}}
\newcommand{\INSTFH}{\affiliation{Duke University, Department of Physics, Durham, North Carolina, U.S.A.}}
\newcommand{\INSTBA}{\affiliation{Ecole Polytechnique, IN2P3-CNRS, Laboratoire Leprince-Ringuet, Palaiseau, France }}
\newcommand{\INSTEF}{\affiliation{ETH Zurich, Institute for Particle Physics, Zurich, Switzerland}}
\newcommand{\INSTEG}{\affiliation{University of Geneva, Section de Physique, DPNC, Geneva, Switzerland}}
\newcommand{\INSTHJ}{\affiliation{University of Glasgow, School of Physics and Astronomy, Glasgow, United Kingdom}}
\newcommand{\INSTDG}{\affiliation{H. Niewodniczanski Institute of Nuclear Physics PAN, Cracow, Poland}}
\newcommand{\INSTCB}{\affiliation{High Energy Accelerator Research Organization (KEK), Tsukuba, Ibaraki, Japan}}
\newcommand{\INSTIB}{\affiliation{University of Houston, Department of Physics, Houston, Texas, U.S.A.}}
\newcommand{\INSTED}{\affiliation{Institut de Fisica d'Altes Energies (IFAE), The Barcelona Institute of Science and Technology, Campus UAB, Bellaterra (Barcelona) Spain}}
\newcommand{\INSTEC}{\affiliation{IFIC (CSIC \& University of Valencia), Valencia, Spain}}
\newcommand{\INSTHH}{\affiliation{Institute For Interdisciplinary Research in Science and Education (IFIRSE), ICISE, Quy Nhon, Vietnam}}
\newcommand{\INSTEI}{\affiliation{Imperial College London, Department of Physics, London, United Kingdom}}
\newcommand{\INSTGF}{\affiliation{INFN Sezione di Bari and Universit\`a e Politecnico di Bari, Dipartimento Interuniversitario di Fisica, Bari, Italy}}
\newcommand{\INSTBE}{\affiliation{INFN Sezione di Napoli and Universit\`a di Napoli, Dipartimento di Fisica, Napoli, Italy}}
\newcommand{\INSTBF}{\affiliation{INFN Sezione di Padova and Universit\`a di Padova, Dipartimento di Fisica, Padova, Italy}}
\newcommand{\INSTBD}{\affiliation{INFN Sezione di Roma and Universit\`a di Roma ``La Sapienza'', Roma, Italy}}
\newcommand{\INSTEB}{\affiliation{Institute for Nuclear Research of the Russian Academy of Sciences, Moscow, Russia}}
\newcommand{\INSTHI}{\affiliation{Institute of Physics (IOP), Vietnam Academy of Science and Technology (VAST), Hanoi, Vietnam}}
\newcommand{\INSTHA}{\affiliation{Kavli Institute for the Physics and Mathematics of the Universe (WPI), The University of Tokyo Institutes for Advanced Study, University of Tokyo, Kashiwa, Chiba, Japan}}
\newcommand{\INSTCC}{\affiliation{Kobe University, Kobe, Japan}}
\newcommand{\INSTCD}{\affiliation{Kyoto University, Department of Physics, Kyoto, Japan}}
\newcommand{\INSTEJ}{\affiliation{Lancaster University, Physics Department, Lancaster, United Kingdom}}
\newcommand{\INSTFC}{\affiliation{University of Liverpool, Department of Physics, Liverpool, United Kingdom}}
\newcommand{\INSTFI}{\affiliation{Louisiana State University, Department of Physics and Astronomy, Baton Rouge, Louisiana, U.S.A.}}
\newcommand{\INSTHB}{\affiliation{Michigan State University, Department of Physics and Astronomy,  East Lansing, Michigan, U.S.A.}}
\newcommand{\INSTCE}{\affiliation{Miyagi University of Education, Department of Physics, Sendai, Japan}}
\newcommand{\INSTDF}{\affiliation{National Centre for Nuclear Research, Warsaw, Poland}}
\newcommand{\INSTFJ}{\affiliation{State University of New York at Stony Brook, Department of Physics and Astronomy, Stony Brook, New York, U.S.A.}}
\newcommand{\INSTGJ}{\affiliation{Okayama University, Department of Physics, Okayama, Japan}}
\newcommand{\INSTCF}{\affiliation{Osaka City University, Department of Physics, Osaka, Japan}}
\newcommand{\INSTGG}{\affiliation{Oxford University, Department of Physics, Oxford, United Kingdom}}
\newcommand{\INSTGC}{\affiliation{University of Pittsburgh, Department of Physics and Astronomy, Pittsburgh, Pennsylvania, U.S.A.}}
\newcommand{\INSTFA}{\affiliation{Queen Mary University of London, School of Physics and Astronomy, London, United Kingdom}}
\newcommand{\INSTE}{\affiliation{University of Regina, Department of Physics, Regina, Saskatchewan, Canada}}
\newcommand{\INSTGD}{\affiliation{University of Rochester, Department of Physics and Astronomy, Rochester, New York, U.S.A.}}
\newcommand{\INSTHC}{\affiliation{Royal Holloway University of London, Department of Physics, Egham, Surrey, United Kingdom}}
\newcommand{\INSTBC}{\affiliation{RWTH Aachen University, III. Physikalisches Institut, Aachen, Germany}}
\newcommand{\INSTFB}{\affiliation{University of Sheffield, Department of Physics and Astronomy, Sheffield, United Kingdom}}
\newcommand{\INSTDI}{\affiliation{University of Silesia, Institute of Physics, Katowice, Poland}}
\newcommand{\INSTIA}{\affiliation{SLAC National Accelerator Laboratory, Stanford University, Menlo Park, California, USA}}
\newcommand{\INSTBB}{\affiliation{Sorbonne Universit\'e, Universit\'e Paris Diderot, CNRS/IN2P3, Laboratoire de Physique Nucl\'eaire et de Hautes Energies (LPNHE), Paris, France}}
\newcommand{\INSTEH}{\affiliation{STFC, Rutherford Appleton Laboratory, Harwell Oxford,  and  Daresbury Laboratory, Warrington, United Kingdom}}
\newcommand{\INSTCH}{\affiliation{University of Tokyo, Department of Physics, Tokyo, Japan}}
\newcommand{\INSTBJ}{\affiliation{University of Tokyo, Institute for Cosmic Ray Research, Kamioka Observatory, Kamioka, Japan}}
\newcommand{\INSTCG}{\affiliation{University of Tokyo, Institute for Cosmic Ray Research, Research Center for Cosmic Neutrinos, Kashiwa, Japan}}
\newcommand{\INSTHF}{\affiliation{Tokyo Institute of Technology, Department of Physics, Tokyo, Japan}}
\newcommand{\INSTGI}{\affiliation{Tokyo Metropolitan University, Department of Physics, Tokyo, Japan}}
\newcommand{\INSTHG}{\affiliation{Tokyo University of Science, Faculty of Science and Technology, Department of Physics, Noda, Chiba, Japan}}
\newcommand{\INSTF}{\affiliation{University of Toronto, Department of Physics, Toronto, Ontario, Canada}}
\newcommand{\INSTB}{\affiliation{TRIUMF, Vancouver, British Columbia, Canada}}
\newcommand{\INSTG}{\affiliation{University of Victoria, Department of Physics and Astronomy, Victoria, British Columbia, Canada}}
\newcommand{\INSTDJ}{\affiliation{University of Warsaw, Faculty of Physics, Warsaw, Poland}}
\newcommand{\INSTDH}{\affiliation{Warsaw University of Technology, Institute of Radioelectronics, Warsaw, Poland}}
\newcommand{\INSTFD}{\affiliation{University of Warwick, Department of Physics, Coventry, United Kingdom}}
\newcommand{\INSTGH}{\affiliation{University of Winnipeg, Department of Physics, Winnipeg, Manitoba, Canada}}
\newcommand{\INSTEA}{\affiliation{Wroclaw University, Faculty of Physics and Astronomy, Wroclaw, Poland}}
\newcommand{\INSTHE}{\affiliation{Yokohama National University, Faculty of Engineering, Yokohama, Japan}}
\newcommand{\INSTH}{\affiliation{York University, Department of Physics and Astronomy, Toronto, Ontario, Canada}}

\INSTHD
\INSTEE
\INSTFE
\INSTD
\INSTGA
\INSTI
\INSTGB
\INSTFG
\INSTFH
\INSTBA
\INSTEF
\INSTEG
\INSTHJ
\INSTDG
\INSTCB
\INSTIB
\INSTED
\INSTEC
\INSTHH
\INSTEI
\INSTGF
\INSTBE
\INSTBF
\INSTBD
\INSTEB
\INSTHI
\INSTHA
\INSTCC
\INSTCD
\INSTEJ
\INSTFC
\INSTFI
\INSTHB
\INSTCE
\INSTDF
\INSTFJ
\INSTGJ
\INSTCF
\INSTGG
\INSTGC
\INSTFA
\INSTE
\INSTGD
\INSTHC
\INSTBC
\INSTFB
\INSTDI
\INSTIA
\INSTBB
\INSTEH
\INSTCH
\INSTBJ
\INSTCG
\INSTHF
\INSTGI
\INSTHG
\INSTF
\INSTB
\INSTG
\INSTDJ
\INSTDH
\INSTFD
\INSTGH
\INSTEA
\INSTHE
\INSTH

\author{K.\,Abe}\INSTBJ
\author{R.\,Akutsu}\INSTCG
\author{A.\,Ali}\INSTBF
\author{C.\,Andreopoulos}\INSTEH\INSTFC
\author{L.\,Anthony}\INSTFC
\author{M.\,Antonova}\INSTEC
\author{S.\,Aoki}\INSTCC
\author{A.\,Ariga}\INSTEE
\author{Y.\,Ashida}\INSTCD
\author{Y.\,Awataguchi}\INSTGI
% ...

\collaboration{The T2K Collaboration}\noaffiliation
\date{\today}

\begin{abstract}\newpage
This paper reports on the search for heavy neutrinos with masses in the range $140 < M_N < 493$~MeV/c$^2$ using the off-axis near detector ND280 of the T2K experiment. These particles can be produced from kaon decays in the standard neutrino beam and then subsequently decay in ND280. The decay modes under consideration are $N \to \ell^{\pm}_{\alpha} \pi^{\mp}$ and $N \to \ell^+_{\alpha} \ell^-_{\beta} \brabar\nu$ ($\alpha,\beta=e,\mu$). A search for such events has been made using the Time Projection Chambers of ND280, where the background has been reduced to less than two events in the current dataset in all channels. No excess has been observed in the signal region. A combined Bayesian statistical approach has been applied to extract upper limits on the mixing elements of heavy neutrinos to electron-, muon- and tau- flavoured currents ($U_e^2$, $U_{\mu}^2$, $U_{\tau}^2$) as a function of the heavy neutrino mass, e.g. $U_e^2 < 10^{-9}$ at $90\%$ C.L. for a mass of $390$\,MeV/c$^2$. These constraints are competitive with previous experiments.
\end{abstract}

\maketitle

\section{Introduction}
\label{sec:introduction}

Neutrino oscillations provide strong evidence that neutrinos are massive particles. Although in the minimal Standard Model they are massless, the most natural extension to allow non-zero masses compatible with oscillation experiments results (two different $\Delta m^2$) consists in the introduction of $n \geq 2$ new right-handed (sterile) neutrino fields $\nu_R$ with the following mass term \cite{hNu:pheno}:

\begin{align}\label{eq:lagrangian}
\mathcal{L}_{\text{mass}} = -\frac{1}{2} 
\left( \begin{array}{cc} \bar\nu_L\ \bar\nu_R^c \end{array} \right) 
\left( \begin{array}{cc} 0 & m_D \\ m_D^T & m_R \end{array} \right)
\left( \begin{array}{c} \nu_L^c \\ \nu_R \end{array} \right) + \hc,
\end{align}
where $m_D$ is the $3 \times n$ Dirac mass matrix and $m_R$ is the $n \times n$ Majorana mass matrix. If the seesaw condition $m_D^T m_D \ll m_R^2$ holds (in terms of eigenvalues), diagonalisation of the mass matrix yields three light Majorana mass eigenstates $\nu_{i}$ ($i=1,2,3$), with masses $m_{\nu,i}$ of the order of the eigenvalues of $m_D m_R^{-1} m_D^T$ and $n$ heavy Majorana mass eigenstates $N_{I}$ ($I=1,\ldots,n$) (heavy neutrinos, also called heavy neutral leptons in the literature), with masses $M_{N,I}$ of the order of the eigenvalues of $m_R$. The flavour eigenstates can be expressed in terms of the mass eigenstates as:

\begin{equation}\label{eq:interaction_eigenstate}
\nu_{\alpha} = \sum_{i=1}^3 V_{\alpha i} \nu_i + \sum_{I=1}^n \Theta_{\alpha I} N_I \quad (\alpha = e,\mu,\tau),
\end{equation}
where $V$ corresponds to the usual PMNS matrix and $\Theta$ is the active-heavy mixing matrix. Heavy neutrinos can be produced in leptonic meson decays $M^{\pm} \to \ell_{\alpha}^{\pm} + N_I$ with a branching ratio proportional to $\vert \Theta_{\alpha I} \vert^2$, for $M_N < m_{\text{meson}} - m_{\ell_{\alpha}}$ and they can similarly decay via the same mixing element.

If at least two of the heavy neutrinos have a mass between $0.1$ and $100$ GeV/c$^2$, they can generate baryogenesis via leptogenesis without any additional new physics~\cite{hNu:leptogenesis}. An example of such a model is the Neutrino Minimal Standard Model ($\nu$MSM) with $n=3$, in which $N_1$ has a mass of $1-100$ keV/c$^2$ and is a warm dark matter candidate, while $N_{2,3}$ are degenerate with GeV-scale masses~\cite{hNu:nuMSM1,hNu:nuMSM2}.

In the following, we define $U_{\alpha}^2 \equiv \sum \vert \Theta_{\alpha I} \vert^2$ summing over the heavy neutrinos that cannot be distinguished experimentally (such as $N_2$ and $N_3$ in the $\nu$MSM).

Limits on $U_{\alpha}^2$ for $M_N < 493$ MeV/c$^2$ can be obtained either by studying heavy neutrino production from kaon decays ($K^{\pm} \to \ell^{\pm} N$) or by searching for heavy neutrino decays, e.g. to one pion and one charged lepton ($N \to \ell_{\alpha}^{\pm} \pi^{\mp}$). The best constraints in this mass range were obtained by the BNL E949~\cite{hNu:E949} and the CERN PS191~\cite{hNu:PS191-1,hNu:PS191-2} experiments with limits of the order of $10^{-9} - 10^{-8}$ on $U_e^2$, $U_{\mu}^2$ and $U_e U_{\mu}$ for $M_N = 200-450$ MeV/c$^2$. Limits from other experiments are summarised in the review \cite{hNu:limits}.

This paper presents the search for potential heavy neutrinos produced in the T2K decay volume and decaying in the T2K Near Detector, ND280, as was originally suggested in~\cite{hNu:Asaka}.

\section{Experimental setup}
\label{sec:experiment}

\subsection{The T2K beamline}
\label{sec:T2K}

The Tokai-to-Kamioka (T2K) experiment~\cite{NIM:T2K} is a long-baseline neutrino experiment located in Japan with the primary goal of measuring muon (anti-)neutrino oscillations using Super-Kamiokande as its far detector. The T2K neutrino beam is produced at the Japan Proton Accelerator Research Complex (J-PARC) by colliding $30$\,GeV protons on a graphite target. The pions and kaons produced are focused and selected by charge with magnetic horns and subsequently decay in flight to neutrinos. Depending on the polarity of the current in the horns, the experiment can be run either in neutrino or anti-neutrino mode.

In this analysis, the production of heavy neutrinos from kaon decays in data taken from November 2010 to May 2017 are considered. This corresponds to a total exposure of $12.34 \times 10^{20}$ protons-on-target (POT) in neutrino mode and $6.29 \times 10^{20}$ POT in anti-neutrino mode, after data quality cuts.

\subsection{The off-axis near detector ND280}
\label{sec:ND280}

\begin{figure}[hbtp]
\centering
\includegraphics[width=0.9\linewidth]{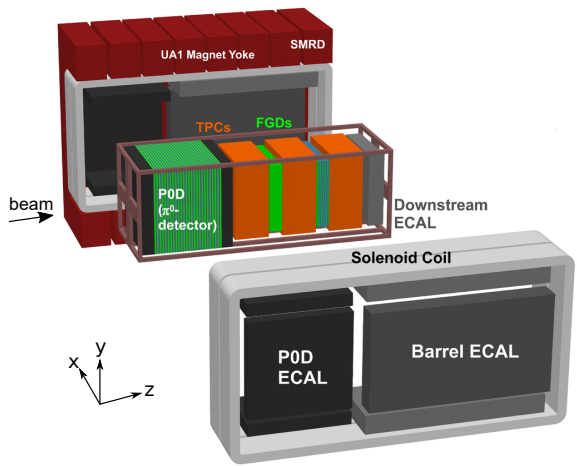}
\caption{An exploded view of the ND280 off-axis near detector
labelling each sub-detector. Adapted from~\cite{NIM:T2K}.}
\label{fig:ND280}
\end{figure}

The off-axis near detector ND280 is located 280 metres from the proton target. It is composed of several sub-detectors with a $0.2$\,T magnet~\cite{NIM:T2K}. The central tracker consists of three Time Projection Chambers (TPCs)~\cite{NIM:TPC}, two scintillator-based Fine-Grained Detectors (FGDs)~\cite{NIM:FGD} and one $\pi^0$ detector (P0D). It is surrounded by an Electromagnetic Calorimeter (ECal) and a Side Muon Range Detector (SMRD). A schematic view of ND280 is shown in \autoref{fig:ND280}.

The main goal of ND280 is to detect neutrino interactions in order to constrain both neutrino flux and cross-section parameters.
The TPCs are filled with a gas mixture based on argon gas and provide excellent track and momentum reconstruction with a typical resolution of 8\% for 1 GeV/c tracks \cite{T2K:numu}. This can be combined with energy loss (dE/dx) measurements in order to perform particle identification (PID) of charged tracks crossing the TPCs.

The analysis focuses on heavy neutrino decays occurring in the ND280 TPC gas volumes, which corresponds to a total volume of interest of $6.3$\,m$^3$.

\section{Analysis}
\label{sec:analysis}

\subsection{Simulation}
\label{sec:simulation}

\begin{figure}[hbtp]
\centering
\includegraphics[width=\linewidth]{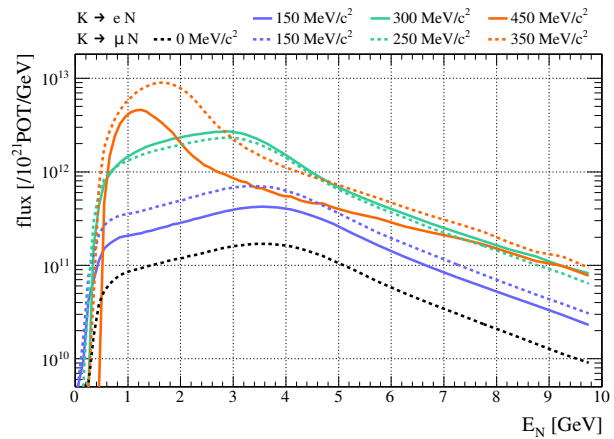}
\caption{Expected flux of heavy neutrinos crossing the ND280 TPCs from $K^{\pm} \to \mu^{\pm} N$ and $K^{\pm} \to e^{\pm} N$ for several values of $M_N$, with the T2K beam in neutrino-mode and for $U_e^2 = U_{\mu}^2 = 1$. \changes{The black dotted curve corresponds to the limiting case of a massless neutrino ($N=\nu$); the one for $K\to e \nu$ is not drawn as it is a few orders of magnitude lower due to helicity suppression.}}
\label{fig:flux}
\end{figure}

The simulation of heavy neutrino production and decay is performed using the T2K neutrino flux prediction, which is constrained by the NA61/SHINE experiment results and by in-situ measurements~\cite{t2k:flux,t2k:NA61}.
We first consider the flux of standard light neutrinos coming from kaon decays in the beamline and crossing the ND280 TPCs. This flux is transformed into a flux of heavy neutrinos ($K^{\pm} \to \ell^{\pm}_{\alpha} N$, $\alpha=e,\mu$) by weighting event-by-event using the appropriate branching ratios~\cite{hNu:Shrock1,hNu:Shrock2,hNu:branching} and modified kinematics. The analysis assumes the heavy neutrino lifetime is long enough to reach ND280 \changes{($\tau \gg \text{time of flight} \sim 1\,\mu$s)}, which is consistent with current limits on the mixing elements. \autoref{fig:flux} presents the results of the simulation for different heavy neutrino masses and for both production modes in neutrino mode. The flux has the same shape for anti-neutrino mode, although it is a factor of $\sim 3$ lower.

\begin{figure}[hbtp]
\centering
\includegraphics[width=\linewidth]{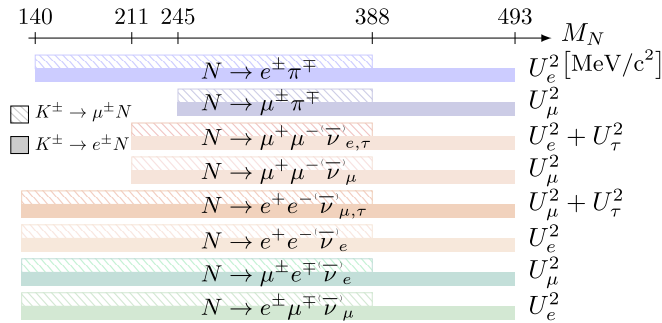}
\caption{Schematic of all the possible production and decay mode combinations for heavy neutrinos with $140<M_N<493$ MeV/c$^2$. The coloured bars show the allowed kinematic regions for each decay mode, with the corresponding mixing element in the right column. A total of 40 production/decay mode combinations are possible.}
\label{fig:HNL_prod}
\end{figure}

The heavy neutrino decays are then simulated at a random point along their trajectories inside ND280. All the possible modes $N\to\ell^{\pm}\pi^{\mp}$ and $N\to\ell^{\pm}\ell^{\mp}\brabar\nu$ were simulated. \autoref{fig:HNL_prod} shows the allowed production and decay modes as a function of the heavy neutrino mass. The neutral current decay modes $N \to e^+ e^- \brabar\nu_{\tau}$ and $N \to \mu^+ \mu^- \brabar\nu_{\tau}$ are directly sensitive to the mixing element $U_{\tau}^2$.

Effects related to heavy neutrino polarisation~\cite{hNu:helicity} and delayed arrival time (with respect to light neutrinos) are taken into account in the simulation.

\subsection{Selection}
\label{sec:selection}

The selection was developed to isolate the signal events listed in \autoref{fig:HNL_prod} from the background expected from standard neutrino interactions with matter. In order to significantly improve the signal to background ratio, which is inversely proportional to the density of the medium, only events occurring in the TPC gas volume are considered for this analysis. 

Events are pre-selected by identifying two tracks of opposite charge originating from a vertex in a TPC. There should be no other tracks in the TPC itself or in the detector located directly upstream (e.g. P0D for the first TPC or the first FGD for the second TPC). Particle identification for each individual track is performed using energy loss in the TPC. Five channels are then identified: $\mu^{\pm}\pi^{\mp}$, $e^- \pi^+$, $e^+ \pi^-$, $e^+ e^-$, $\mu^+ \mu^-$. 

In the analysis, we do not define any specific selection for the three-body decays $N \to e^{\pm} \mu^{\mp} \nu$ because these modes already contribute to the $e^{\pm} \pi^{\mp}$ selection channels. For the $\mu^+ \mu^-$ channel, electromagnetic calorimeter information is also used to clearly identify the two muons.

Several kinematic cuts are then applied to further reject the background:
\begin{itemize}
\item invariant mass $m_{\rm inv}$ of the two-track system: in the case of a heavy neutrino decay, it is expected that $m_{\rm inv}^{\rm true} \leq M_N$ ($m_{\rm inv}^{\rm true} = M_N$ for the two-body decays). The heavy neutrino is produced in kaon decays so that it is necessarily lighter than $M_K = 493$ MeV/c$^2$ allowing an upper cut on the reconstructed invariant mass $m_{\rm inv}^{\rm reco} < 700$\,MeV/c$^2$ to be applied. The additional margin accounts for detector resolution effects.
\item angle between the two tracks $\Delta\Phi$: the two charged tracks produced in the decay are boosted forward so that only events with $\Delta\Phi < 90\degree$ can be selected without loss of signal efficiency.
\item incoming heavy neutrino polar angle $\theta$: the heavy neutrino's direction is collinear to the beam, while the products of an active neutrino interaction are expected to be distributed with a larger angle because of potential missed tracks or nuclear effects. $\theta$ is reconstructed using the two charged tracks but can still be used with a good approximation for the three-body decays. The cut is $\cos\theta > 0.992$ for $\mu^{\pm}\pi^{\mp}$ for the $\mu\pi$ channel and $\cos\theta > 0.99$ for the others. 
\end{itemize}

\begin{figure}[hbtp]
\includegraphics[width=\linewidth]{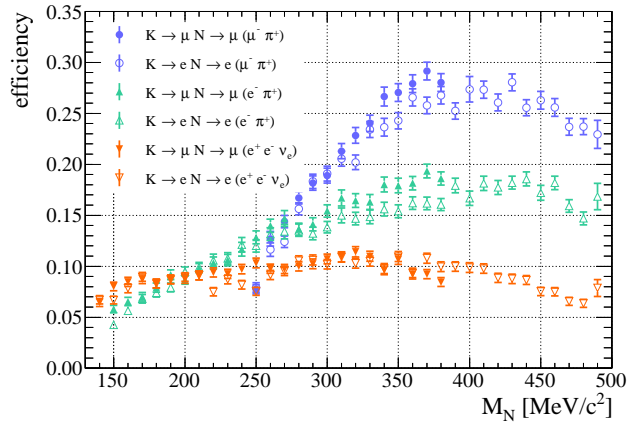}
\caption{Heavy neutrino signal selection efficiency in neutrino mode as a function of heavy neutrino mass for some of the decay modes. Error bars include both statistical and systematic uncertainties.}
\label{fig:eff_FHC}
\end{figure}

Applying these criteria to the signal simulated in the ND280 TPC gas volumes, the efficiencies of the signal selection for the different modes were obtained, as shown in~\autoref{fig:eff_FHC}. For a given mass, they are quite independent of the production mode ($K^{\pm} \to \mu^{\pm} N$ or $K^{\pm} \to e^{\pm} N$). $\mu\pi$ efficiencies are slightly better as muon tracks are easier to reconstruct in the TPC.

\subsection{Signal systematic uncertainties}
\label{sec:systematics}

Two sources of systematic uncertainties on the heavy neutrino signal are considered:

\begin{itemize}
\item flux: uncertainties on the kaon flux used as input to the simulation, as presented in \autoref{sec:simulation}, are directly transposed into uncertainties on the flux of heavy neutrinos reaching ND280. The total normalisation uncertainty has been estimated to be $15\%$, using external data such as those from the NA61/SHINE experiment~\cite{t2k:flux}.
\item signal selection efficiency: detector systematic uncertainties are defined to cope with any discrepancies between data and simulation of the detector effects. The dominant uncertainties are related to TPC reconstruction and particle identification performances and have been computed as in previous ND280 analyses~\cite{T2K:xsec}. The overall effects have been estimated to be approximately $5\%$.
\end{itemize}

\subsection{Background estimation}
\label{sec:background}

The background remaining after the selection has first been estimated using the NEUT 5.3.2 Monte Carlo generator \cite{NEUT}, before being constrained using control regions in ND280 data.

One of the dominant background contributions is the neutrino-induced coherent pion production on argon nuclei in the TPC gas ($\nu_{\mu} + \text{Ar} \to \mu^- + \pi^+ + \text{Ar}$). The NEUT prediction has been tuned to T2K and MINERvA data~\cite{bkg:coh_T2K,bkg:coh_MINERvA} with a $30\%$ normalisation uncertainty.

Additional background sources include other types of neutrino interactions in the gas and interactions outside the gas. An example of the latter is the conversion of a photon, emitted by a neutrino interaction in a FGD, to an electron-positron pair. 

Data and simulations are compared with two sets of control regions in order to estimate the model uncertainty on the background. First, a selection of events similar to the signal events, but where the kinematic cut on the polar angle $\theta$ is inverted (CR-I), contains mostly resonant pion production and quasi-elastic processes on argon. Similarly, control regions are identified by considering events starting in the borders of the TPC (meaning the box containing the gas) as the volume of interest rather than in the gas itself (CR-II). These control regions are dominated by photon conversions and other mis-reconstructed processes.

\autoref{tab:CS} presents the comparison of T2K data and NEUT predictions in the aforementioned control regions. We have not found any significant discrepancies between data and Monte Carlo predictions in any of them. Conservatively, we have assigned to each background source a model uncertainty equal to the statistical uncertainty of the data in the corresponding control region. 

\begin{table}
\caption{Comparison of number of events in data (D) and corresponding NEUT prediction (with statistical uncertainties) in the control regions used to determine the model uncertainties in the different channels, using the data set presented in \autoref{sec:T2K}.}
\label{tab:CS}

\squeezetable
\renewcommand*{\arraystretch}{1.2}
\begin{tabular}{c|cc|cc|cc|cc}
\hline \hline
 & \multicolumn{4}{c|}{Neutrino mode} & \multicolumn{4}{c}{Anti-neutrino mode} \\
 \cline{2-9}
 & \multicolumn{2}{c|}{CR-I} & \multicolumn{2}{c|}{CR-II} & \multicolumn{2}{c|}{CR-I} & \multicolumn{2}{c}{CR-II} \\
\textbf{Ch.} & D & NEUT & D & NEUT & D & NEUT & D & NEUT\\
\hline
$\mu^{\pm} \pi^{\mp}$ & $15$ & $11.4 \pm 1.0$ & $36$ & $30.1 \pm 1.6$ & $2$ & $2.6 \pm 0.5$ & $5$ & $6.5 \pm 0.8$ \\
$e^{-} \pi^{+}$ & $2$ & $2.4 \pm 0.5$ & $14$ & $13.3 \pm 1.1$ & $2$ & $0.7 \pm 0.2$ & $7$ & $5.4 \pm 0.7$ \\
$e^{+} \pi^{-}$ & $5$ & $3.1 \pm 0.5$ & $17$ & $11.7 \pm 1.0$ & $1$ & $0.5 \pm 0.2$ & $2$ & $3.3 \pm 0.6$ \\
$\mu^{+} \mu^{-}$ & $0$ & $0.9 \pm 0.3$ & $3$ & $2.5 \pm 0.5$ & $0$ & $0.2 \pm 0.1$ & $1$ & $0.4 \pm 0.2$ \\
$e^{+} e^{-}$ & $4$ & $4.5 \pm 0.6$ & $58$ & $53.5 \pm 2.2$ & $3$ & $0.8 \pm 0.3$ & $13$ & $14.3 \pm 1.3$ \\
\hline \hline
\end{tabular}
\end{table}

For a given channel, the number of expected background events is the nominal value from NEUT and the total uncertainty is the sum of the contributions from the Monte Carlo statistical uncertainty, the flux and detector systematic uncertainties and the model uncertainties described above. 

\begin{table}[hbtp]
\caption{Summary of the estimated numbers of background events in the different analysis channels in neutrino and anti-neutrino beam modes with the corresponding absolute uncertainties (statistical, flux-related, detector-related, model), using the data set presented in \autoref{sec:T2K}.}
\label{tab:bkg}

\centering
\renewcommand*{\arraystretch}{1.2}
\begin{tabular}{c|c|c|cccc|c}
\hline \hline
\multirow{2}{*}{\textbf{Mode}} & \multirow{2}{*}{\textbf{Ch.}} & Expected & \multicolumn{5}{c}{Uncertainties} \\
\cline{4-8}
& & background & stat. & flux & det. & model & total \\
\hline
\multirow{5}{*}{\rotatebox[origin=c]{90}{\parbox[c]{1cm}{\centering \small neutrino}}}
& $\mu^{\pm} \pi^{\mp}$ & 1.543 & 0.366 & 0.154 & 0.165 & 0.285 & 0.516 \\
& $e^{-} \pi^{+}$ & 0.376 & 0.213 & 0.038 & 0.104 & 0.097 & 0.259 \\
& $e^{+} \pi^{-}$ & 0.328 & 0.186 & 0.033 & 0.117 & 0.115 & 0.250 \\
& $\mu^{+} \mu^{-}$ & 0.216 & 0.107 & 0.022 & 0.045 & 0.062 & 0.133 \\
& $e^{+} e^{-}$ & 0.563 & 0.192 & 0.056 & 0.092 & 0.074 & 0.233 \\
\hline
\multirow{5}{*}{\rotatebox[origin=c]{90}{\parbox[c]{1cm}{\centering anti-neutrino}}}
& $\mu^{\pm} \pi^{\mp}$ & 0.384 & 0.161 & 0.038 & 0.058 & 0.100 & 0.202 \\
& $e^{-} \pi^{+}$ & 0.018 & 0.018 & 0.002 & 0.005 & 0.005 & 0.020 \\
& $e^{+} \pi^{-}$ & 0.219 & 0.155 & 0.022 & 0.140 & 0.122 & 0.243 \\
& $\mu^{+} \mu^{-}$ & 0.038 & 0.038 & 0.004 & 0.007 & 0.011 & 0.040 \\
& $e^{+} e^{-}$ & 0.015 & 0.015 & 0.002 & 0.001 & 0.004 & 0.016 \\
\hline \hline
\end{tabular} 
\end{table}

\autoref{tab:bkg} summarises the background in the different analysis channels. The dominant contribution to its uncertainty comes from the limited statistics of the samples. The background in the $\mu\pi$ channel is higher than for other channels, as it is dominated by the irreducible coherent pion production.

\subsection{Statistical analysis}
\label{sec:stat}

Two approaches have been considered to constrain the mixing elements $U_e^2$, $U_{\mu}^2$ and $U_{\tau}^2$. 

In the first approach, each heavy neutrino production/decay mode is considered independently and the corresponding analysis channel is used to put limits on the associated mixing elements. For instance, the $\mu^{\pm}\pi^{\mp}$ channel as defined in \autoref{sec:selection} can constrain:

\begin{itemize} 
\item either $U_{\mu}^2$ by considering only the signal from \mbox{$K^{\pm} \to \mu^{\pm} N$, $N \to \mu^{\pm} \pi^{\mp}$},
\item or $U_e \times U_{\mu}$ by considering only the signal from \mbox{$K^{\pm} \to e^{\pm} N$, $N \to \mu^{\pm} \pi^{\mp}$}.
\end{itemize}

Three methods to obtain constraints in this approach have been applied:
\begin{enumerate}
\item[($\mathbf{A}$)] assuming that the background is zero, set conservative upper limits, independently of background modelling and estimation, on the mixing elements using the Highland-Cousin method~\cite{stat:HC};
\item[($\mathbf{B}$)] the Feldman-Cousins method to define confidence intervals, taking into account the non-zero background~\cite{stat:FC};
\item[($\mathbf{C}$)] a Bayesian method to define credible intervals, taking into account the non-zero background.
\end{enumerate}

This ``single-channel" approach has the advantage of being straightforward and is similar to that of the PS191 collaboration~\cite{hNu:PS191-1,hNu:PS191-2}. However, it does not allow the different modes and channels to be combined, so that the constraints are valid only under strong assumptions of the hierarchy of $U_e^2$, $U_{\mu}^2$ and $U_{\tau}^2$. A ``combined" approach was then defined, in which all the heavy neutrino production and decay modes (presented in \autoref{fig:HNL_prod}) and the ten different analysis channels (five for each beam mode) are considered simultaneously. For a given analysis channel $A$, the contribution of a mode $i$ is characterised by:

\begin{itemize}
\item the expected number of decays in the detector assuming $U_e^2 = U_{\mu}^2 = U_{\tau}^2 = 1$ and 100\% selection efficiency, denoted $\Phi_i$ ;
\item the selection efficiency of these decays in the current channel $\varepsilon_{A,i}$ ;
\item the actual values of $U_e^2$, $U_{\mu}^2$ and $U_{\tau}^2$ via the factor $f_i = U_{\alpha}^2 \sum U_{\beta_j}^2$ with $\alpha,\beta_j \in \{e,\mu,\tau\}$ where $\alpha$ is the flavour involved at the production of the heavy neutrino and $\beta_j$ are the flavours involved in its decay (only one for charge current modes, several for neutral current modes).
\end{itemize}

The expected number of events $\mathcal{N}_A$ in a channel $A$ depends on the background in this channel $B_A$ and the sum of the contributions from the different production and decay modes: 

\begin{equation}
\mathcal{N}_A = B_A + \sum_{i} \varepsilon_{A,i} \times f_i(U_e^2, U_{\mu}^2, U_{\tau}^2) \times \Phi_{i}.
\end{equation}

Only a Bayesian method has been considered in this combined approach. The likelihood is built using a Poisson function for the observed number of events $n^{\text{obs}}_{A}$ in each channel $A$, with Poisson parameter $\mathcal{N}_A$:

\begin{equation}
\mathcal{L} = \prod_{A} \text{Poisson} \left( n^{\text{obs}}_{A}, \mathcal{N}_A \right).
\end{equation}

The uncertainties on the flux and efficiency are taken into account in the forms of multivariate Gaussian priors $\pi_{\Phi}$ and $\pi_{\varepsilon}$ respectively. The priors on the background $\pi_B$ are taken to be log-normal with means and standard deviations given by the expected background and its uncertainty in \autoref{tab:bkg}. The priors on the mixing elements $U_{\alpha}^2$ are assumed to be flat.

The marginalised posterior probability $p$ is then defined as the product of the likelihood $\mathcal{L}$ and the priors, integrating over all the nuisance parameters (flux, efficiency and background):

\begin{equation}
p(U_e^2, U_{\mu}^2, U_{\tau}^2) = \int d\Phi\,d\varepsilon\,dB \times \mathcal{L} \times \pi_{\Phi}\, \pi_{\varepsilon}\, \pi_{B}\, \pi_{U^2}.
\end{equation}

A Markov Chain Monte Carlo method has been implemented using PyMC~\cite{stat:pymc} to perform this integration. The output can then be used to define 90\% domains, either by profiling or by marginalising over the two other mixing elements. For instance,
\begin{align}
    p_{\text{prof}}(U_e^2) &= p(U_e^2, U_{\mu,\text{max}}^2 U_{\tau,\text{max}}^2), \\
    p_{\text{marg}}(U_e^2) &= \int p(U_e^2, U_{\mu}^2, U_{\tau}^2) dU_{\mu}^2 dU_{\tau}^2, \label{eq:lkl_prof}
\end{align}
where $U_{\mu,\text{max}}^2$ and $U_{\tau,\text{max}}^2$ are the values maximising the likelihood.

Limits in 2D/3D parameter space may be obtained as well. Limits on $U_e^2$ can be computed for $140<M_N<493$ MeV/c$^2$, while limits on $U_{\mu}^2$ and $U_{\tau}^2$ can only be computed for $140<M_N<388$ MeV/c$^2$ due to the kinematic constraints presented in \autoref{fig:HNL_prod}.

\section{Results}
\label{sec:results}

Following the selection from \autoref{sec:selection}, no events were observed in any of the different signal regions, which is consistent with the background-only hypothesis, allowing upper limits on $U_e^2$, $U_{\mu}^2$ and $U_{\tau}^2$ to be placed.

\begin{figure}[hbtp]
\includegraphics[width=\linewidth]{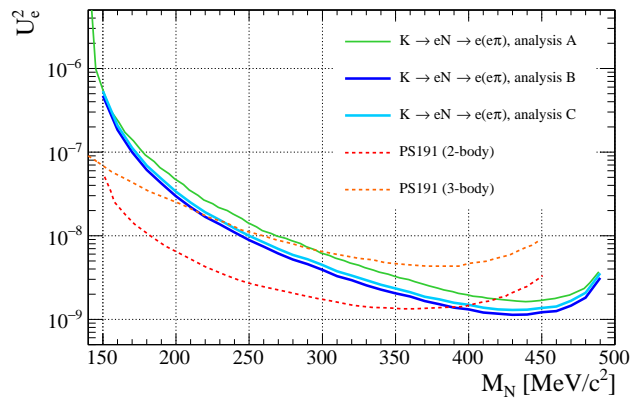}
\caption{90\% upper limits on the mixing element $U_e^2$ as a function of heavy neutrino mass using the single-channel approach, considering only the contribution from $K^{\pm} \to e^{\pm} N, N \to e^{\pm} \pi^{\mp}$, with the three methods $\mathbf{A}$, $\mathbf{B}$ and $\mathbf{C}$.  The limits are compared to the ones of PS191 experiment~\cite{hNu:PS191-1,hNu:PS191-2}.}
\label{fig:results_simple}
\end{figure}

An example of results from the single-channel approach is presented in \autoref{fig:results_simple}. It shows the comparison of the three methods ($\mathbf{A}$, $\mathbf{B}$, $\mathbf{C}$), which give similar upper limits with method $\mathbf{A}$ giving slightly more conservative limits as expected.

\begin{figure}[!h]
\includegraphics[width=\linewidth]{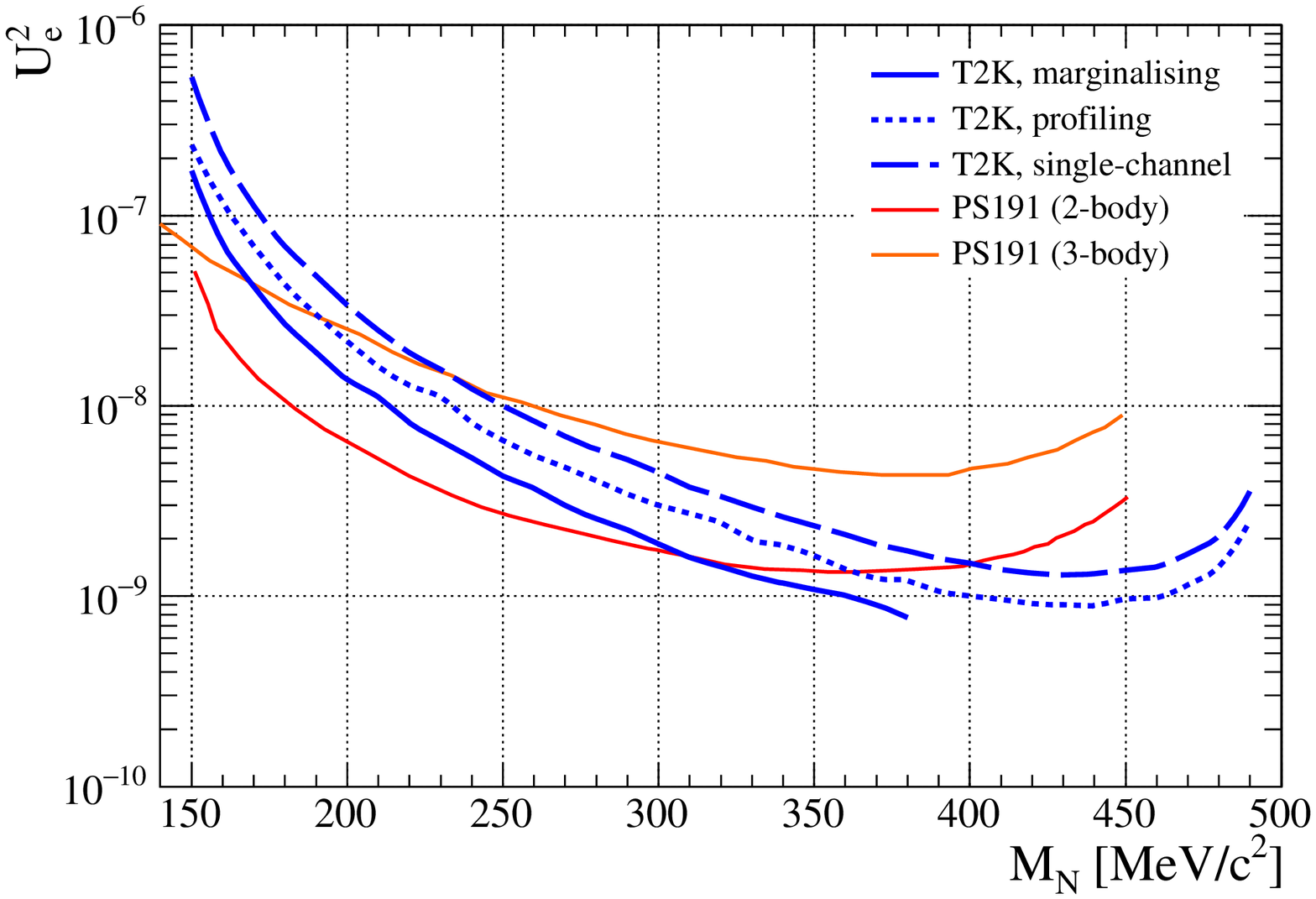}
\includegraphics[width=\linewidth]{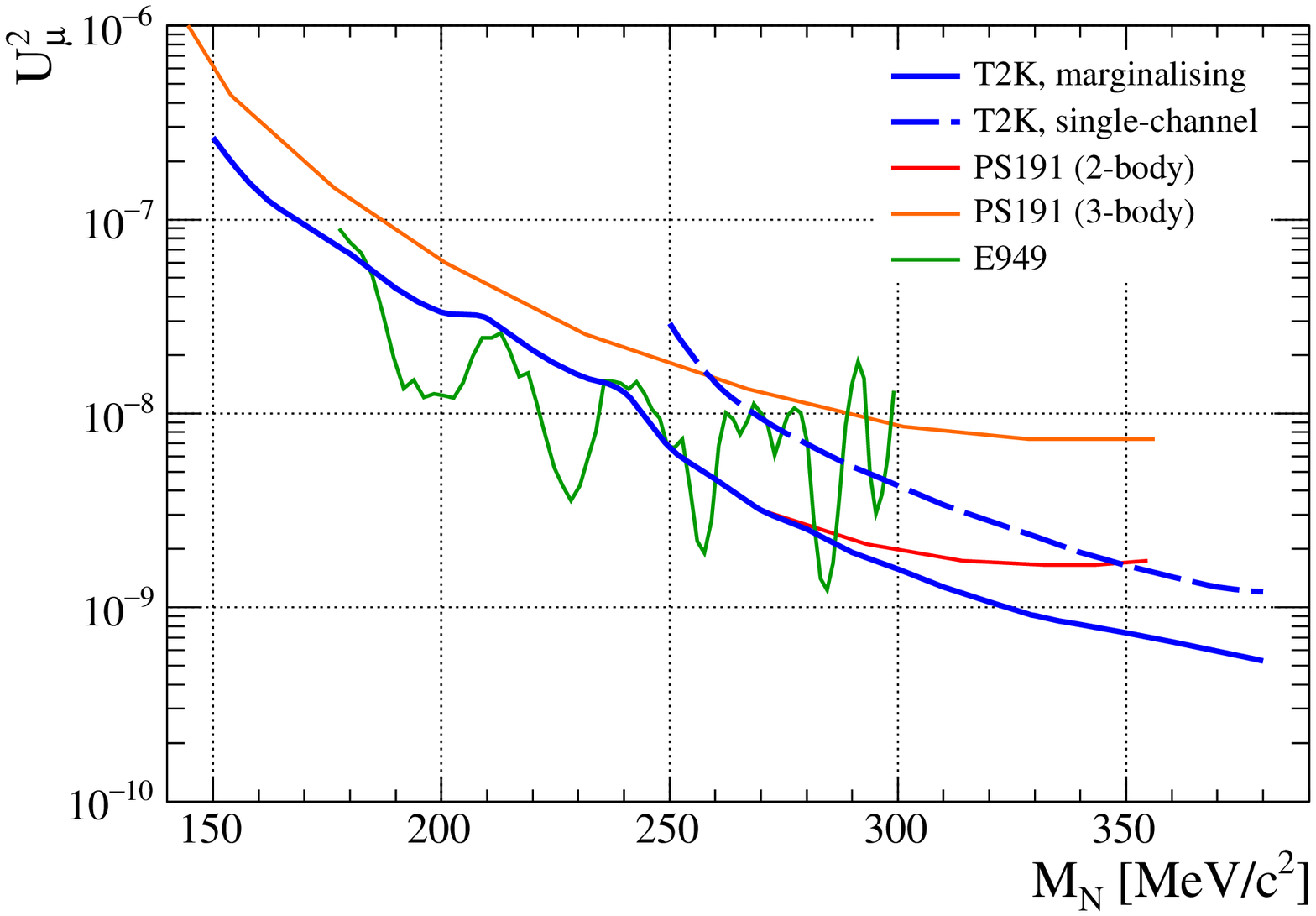}
\includegraphics[width=\linewidth]{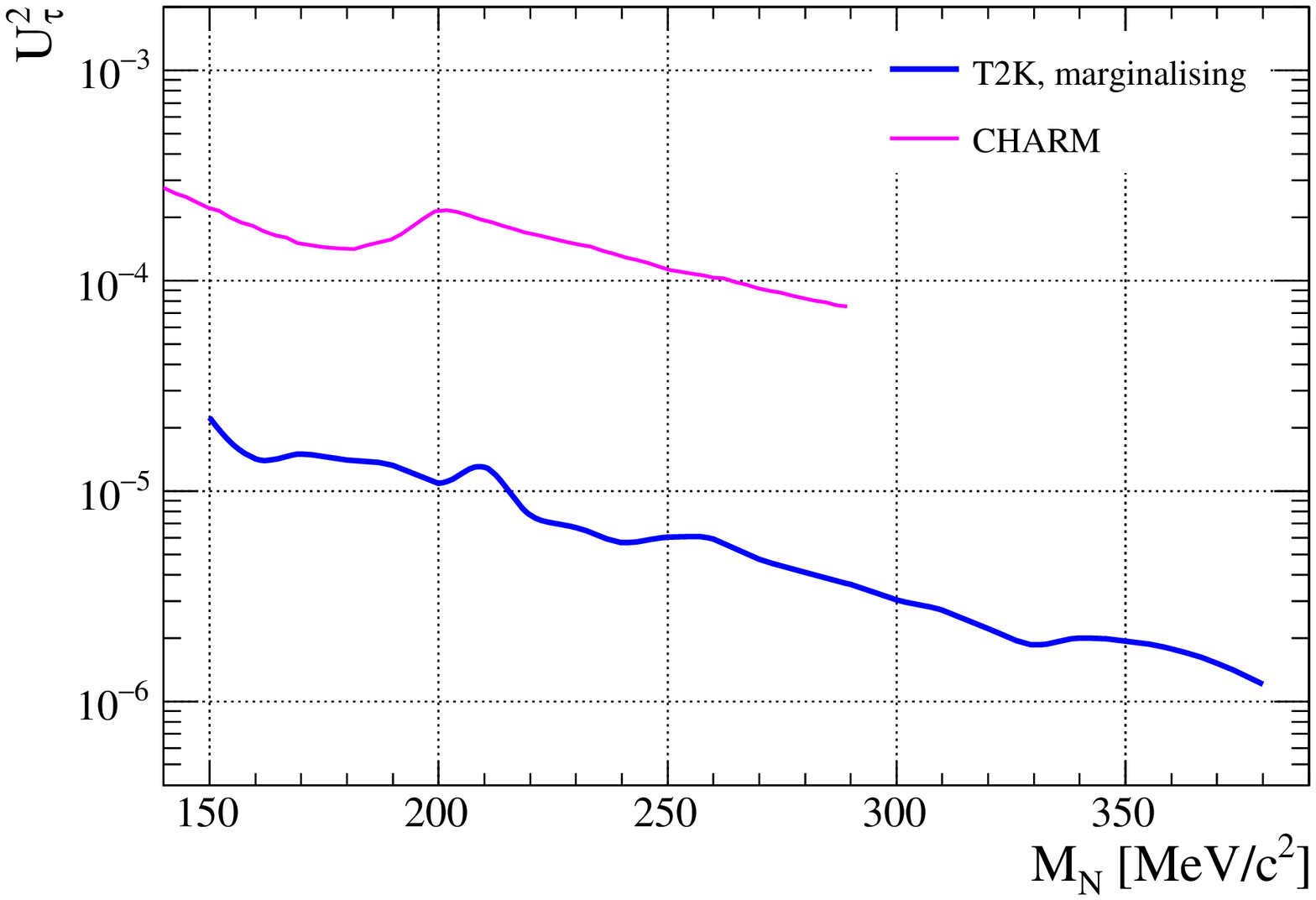}
\caption{90\% upper limits on the mixing elements $U_e^2$ (top), $U_{\mu}^2$ (middle), $U_{\tau}^2$ (bottom) as a function of heavy neutrino mass, obtained with the combined approach. \changes{The blue dashed lines corresponds to the results of the single-channel approach (method C).} The blue solid lines are obtained after marginalisation over the two other mixing elements. In the top plot, the additional blue dotted line corresponds to the case where profiling is used ($U_{\mu}^2=U_{\tau}^2=0$), \changes{as explained in the main text}. The limits are compared to the ones of other experiments: PS191~\cite{hNu:PS191-1,hNu:PS191-2}, E949~\cite{hNu:E949},
CHARM~\cite{hNu:CHARM}.}
\label{fig:results_comb}
\end{figure}

The results of the combined approach are shown in \autoref{fig:results_comb}. 
They provide an improvement by a factor of 2-3 with respect to the single-channel approach, thanks 
to the increased statistical power of the combination.

The limits are competitive with those of previous experiments such as PS191~\cite{hNu:PS191-1,hNu:PS191-2}, E949~\cite{hNu:E949} and CHARM~\cite{hNu:CHARM}, especially in the high-mass region (above $300$ MeV/c$^2$). The kinks clearly visible on $U_{\mu}^2$ and $U_{\tau}^2$ limits come from the changes in the contributing production and decay modes as presented in \autoref{fig:HNL_prod}.

\begin{figure}[!h]
\includegraphics[width=\linewidth]{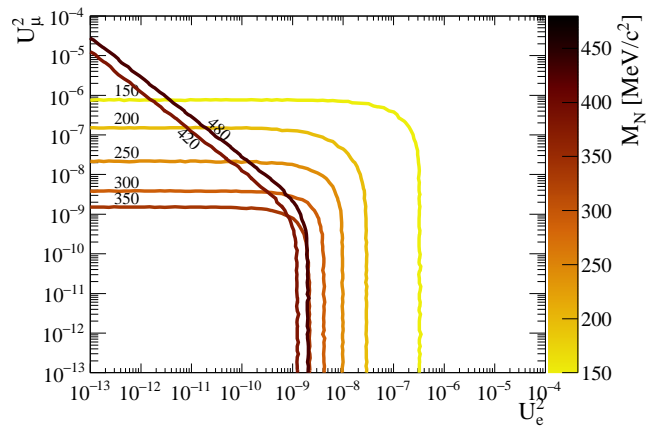}
\caption{2D contours in the $U_e^2 - U_{\mu}^2$ plane, after profiling over $U_{\tau}^2$ ($U_{\tau}^2=0$). Each line corresponds to a different heavy neutrino mass hypothesis.}
\label{fig:results_2D}
\end{figure}

The limits are obtained after marginalisation over the two other mixing elements. For $U_e^2$, the limits after profiling (equation \ref{eq:lkl_prof}) are also presented, which effectively corresponds to setting $U_{\mu}^2 = U_{\tau}^2 = 0$. Indeed, for $M_N > 388$\,MeV/c$^2$, the correlations between $U_e^2$ and $U_{\mu}^2$ (as seen in \autoref{fig:results_2D}) would give limits on $U_e^2$ outside T2K's reach. However, profiling leads to a loss in the sensitivity on $U_e^2$ with respect to the marginalisation as it forcefully suppresses the contributions of the decay modes involving $U_{\mu}^2$ or $U_{\tau}^2$.

It is worth mentioning that the limits depend on the choice of prior on $U_{\alpha}^2$. The limits on $U_e^2$ and $U_{\mu}^2$ are quite robust with respect to a change of prior as T2K data are directly sensitive to these mixing elements (e.g. using $\pi_{U^2}(U_{\alpha}^2) = U_{\alpha}^2$ varies the limit by less than $30\%$), while the limit on $U_{\tau}^2$ is strongly affected (more than $50\%$).

It is also possible to define 2D contours, e.g. in the $U_e^2 - U_{\mu}^2$ plane, allowing the correlations between the mixing elements to be visualised. \autoref{fig:results_2D} presents a set of such contours for different heavy neutrino masses. The change of behaviour at $M_N = 388$\,MeV/c$^2$ corresponds to the kinematic cut-off for $K^{\pm} \to \mu^{\pm} N$ processes as seen in \autoref{fig:HNL_prod}.

\section{Conclusion}
\label{sec:conclusion}

A selection of events with two tracks with opposite charges originating from the ND280 TPC gas volumes allows heavy neutrino decays to be efficiently isolated from expected background coming from standard neutrino interactions with matter. No events are observed in the defined signal regions, which is consistent with the background-only hypothesis.

Limits on the mixing elements $U_e^2$, $U_{\mu}^2$ and $U_{\tau}^2$ are obtained using a combined Bayesian approach. Results apply to any model with heavy neutrinos with masses between $140$\,MeV/c$^2$ and $493$\,MeV/c$^2$ such as \cite{hNu:Rasmussen}, and can, in particular, be interpreted as constraints
on the sum of $N_2$ and $N_3$ coupling squared as explained in the introduction, for the $\nu$MSM.

As the analysis is still statistically limited, results are expected to further improve by a factor of 2-3 with T2K data up to 2026. Additional data will also allow the background treatment to be improved by using more populated control regions. 

By considering heavy neutrino production from pion decays, it would also be possible to extend the phase space down to a few MeV/c$^2$. When combined with a better understanding of the expected background, this may permit the low-mass heavy neutrino phase space ($10<M_N<493$\,MeV/c$^2$ and $U_{e,\mu}^2>10^{-11}-10^{-10}$) to be fully explored.

The results presented in this article are available in the corresponding data release \cite{data_release}. It contains the signal flux and selection efficiencies for all modes and masses, the detailed background predictions, the limits presented in Figures \ref{fig:results_simple}-\ref{fig:results_comb}-\ref{fig:results_2D} and the raw output of the MCMC. One can use the latter to re-compute the limits with different priors on the mixing elements or with different ways to present the results.

\section*{Acknowledgments}

We thank the J-PARC staff for superb accelerator performance. We thank the CERN NA61/SHINE Collaboration for providing valuable particle production data. We acknowledge the support of MEXT, Japan; NSERC (Grant No. SAPPJ-2014-00031), NRC and CFI, Canada; CEA and CNRS/IN2P3, France; DFG, Germany; INFN, Italy; National Science Centre (NCN) and Ministry of Science and Higher Education, Poland; RSF, RFBR, and MES, Russia; MINECO and ERDF funds, Spain; SNSF and SERI, Switzerland; STFC, UK; and DOE, USA. We also thank CERN for the UA1/NOMAD magnet, DESY for the HERA-B magnet mover system, NII for SINET4, the WestGrid and SciNet consortia in Compute Canada, and GridPP in the United Kingdom. In addition, participation of individual researchers and institutions has been further supported by funds from ERC (FP7), ``la Caixa'' Foundation (ID 100010434, fellowship code LCF/BQ/IN17/11620050), the European Union's Horizon 2020 Research and Innovation programme under the Marie Sklodowska-Curie grant agreement no. 713673 and H2020 Grant No. RISE-GA644294-JENNIFER 2020; JSPS, Japan; Royal Society, UK; the Alfred P. Sloan Foundation and the DOE Early Career program, USA; RFBR research project 18-32-00072, Russia.

\bibliography{bibliography}

\end{document}